\documentclass[epj]{webofc}
\usepackage[utf8]{inputenc}
\usepackage[varg]{txfonts}   
\usepackage{booktabs}
\usepackage{xcolor}
\definecolor{darkred}{rgb}{0.4,0.0,0.0}
\definecolor{darkgreen}{rgb}{0.0,0.4,0.0}
\definecolor{darkblue}{rgb}{0.0,0.0,0.4}
\usepackage[bookmarks,linktocpage,colorlinks,
    linkcolor = darkred,
    urlcolor  = darkblue,
    citecolor = darkgreen]{hyperref}
\usepackage{subfigure}
\wocname{EPJ Web of Conferences}
\woctitle{Lattice2017}
\newcommand{\dof}{\mathrm{d.o.f.}}
\newcommand{\HLbL}{\mathrm{HLbL}}

\newcommand{\MsTT}{\overline{\mathcal{M}}_{TT}}
\newcommand{\MsTTa}{\overline{\mathcal{M}}_{TT}^{a}}
\newcommand{\MsTTt}{\overline{\mathcal{M}}_{TT}^{\tau}}
\newcommand{\MsTL}{\overline{\mathcal{M}}_{TL}}
\newcommand{\MsLT}{\overline{\mathcal{M}}_{LT}}
\newcommand{\MsTLt}{\overline{\mathcal{M}}_{TL}^{\tau}}
\newcommand{\MsTLa}{\overline{\mathcal{M}}_{TL}^{a}}
\newcommand{\MsLL}{\overline{\mathcal{M}}_{LL}}

\newcommand{\fm}{\mathrm{fm}}
\newcommand{\MeV}{\mathrm{MeV}}
\newcommand{\GeV}{\mathrm{GeV}}

\newcommand{\phys}{\mathrm{phys}}       
\newcommand{\lat}{\mathrm{lat}}

\newcommand{\GammaGG}{\Gamma_{\gamma\gamma}}

\begin{document}
%
\selectlanguage{english}
\title{%
Light-by-light forward scattering amplitudes in Lattice QCD
}
\author{%
\firstname{Antoine} \lastname{G\'erardin}\inst{1}\fnsep\thanks{Speaker, \email{gerardin@uni-mainz.de} } \and
\firstname{Jeremy} \lastname{Green}\inst{2} \and
\firstname{Oleksii}  \lastname{Gryniuk}\inst{1} \and
\firstname{Georg}  \lastname{von Hippel}\inst{1} \and
\firstname{Harvey B.}  \lastname{Meyer}\inst{1,3} \and
\firstname{Vladimir}  \lastname{Pascalutsa}\inst{1} \and
\firstname{Hartmut}  \lastname{Wittig}\inst{1,3} 
}

\institute{%
Institut f\"ur Kernphysik \& Cluster of Excellence PRISMA,  Johannes Gutenberg-Universit\"at Mainz, D-55099 Mainz, Germany
\and
John von Neumann Institute for Computing (NIC), DESY, Platanenallee 6, 15738 Zeuthen, Germany
\and
Helmholtz Institut Mainz,  D-55099 Mainz, Germany
}
\abstract{%
We present our preliminary results on the calculation of hadronic light-by-light forward scattering amplitudes using vector four-point correlation functions computed on the lattice. Using a dispersive approach, forward scattering amplitudes can be described by $\gamma^* \gamma^* \to$ hadrons fusion cross sections and then compared with phenomenology. We show that only a few states are needed to reproduce our data. In particular, the sum rules considered in this study imply relations between meson$-\gamma\gamma$ couplings 
and provide valuable information about individual form factors which are often used to estimate the meson-pole contributions to the hadronic light-by-light contribution to the $(g-2)$ of the muon.
}
\maketitle
\section{Introduction}\label{intro}

The anomalous magnetic moment of the muon $a_{\mu} = (g-2)_{\mu}/2$ provides a strong test of the Standard Model of particle physics. The persistent $3-4~\sigma$ discrepancy between the experimental result and the theoretical prediction could be a sign of new physics. Two new experiments at Fermilab and J-PARC plan to reduce the error by a factor of four in the next few years~\cite{Venanzoni:2014ixa,Otani:2015lra}. The theory error is dominated by the hadronic vacuum polarisation (HVP) and the Hadronic light-by-light (HLbL) contribution and the latter is expected to dominate in the near future. Two collaborations have presented partial results on the direct calculation of the HLbL contribution \cite{Blum:2014oka,Blum:2015gfa,Blum:2016lnc,Green:2015sra,Green:2015mva,Asmussen:2016lse}. In a complementary way, lattice QCD can provide essential hadronic inputs for the models and dispersive approach to $a_{\mu}^{\HLbL}$~\cite{Colangelo:2014dfa,Colangelo:2015ama,Colangelo:2017qdm,Colangelo:2017fiz,Hagelstein:2017obr}. In particular, a lattice calculation of the pion transition form factor (TFF) can provide the necessary input for the dominant pion-pole contribution~\cite{Gerardin:2016cqj}. The contribution from other narrow pseudoscalar mesons could in principle be computed in a similar way even if the calculation is more demanding due to the presence of large disconnected contributions. However, higher intermediate states are resonances and the direct computation of the associated TFF is much more challenging. 
Two-photon fusion processes are related to the light-by-light forward scattering amplitude via sum rules~\cite{Pascalutsa:2012pr}. Using a phenomenological parametrization of the $\gamma^* \gamma^* \to$ hadron cross sections and fitting the model parameters to the HLbL amplitude computed on the lattice, we are able to test its ability to describe the HLbL amplitude at spacelike virtualities and extract information about single-meson TFFs. The latter can then be used to estimate $a_{\mu}^{\rm HLbL}$ within the model. 

\section{Light-by-light scattering amplitudes and sum rules}\label{sec-1}

The eighty-one light-by-light forward scattering amplitudes $\mathcal{M}_{\lambda_3\lambda_4\lambda_1\lambda_2}$ associated to the process $\gamma^*(\lambda_1,q_1) \ \gamma^*(\lambda_2,q_2) \ \to \ \gamma^*(\lambda_3,q_1) \ \gamma^*(\lambda_4,q_2)$ can be reduced to eight independent amplitudes using parity and time-reversal invariance~\cite{Budnev:1971sz},
\begin{gather}
\nonumber \mathcal{M}_{TT} = \frac{1}{2} (\mathcal{M}_{++,++} + \mathcal{M}_{+-,+-}) \,,\ \mathcal{M}_{TT}^{\tau} = \mathcal{M}_{++,--} \,,\ \mathcal{M}_{TL} =\mathcal{M}_{+0,+0} \,,\ \mathcal{M}_{LT} =M_{0+,0+} \,,\ \mathcal{M}_{LL} =\mathcal{M}_{00,00} \,, \\ 
\mathcal{M}_{TL}^{\tau} = \frac{1}{2} (\mathcal{M}_{++,00} + \mathcal{M}_{0+,-0})  \,, \mathcal{M}_{TT}^{a} = \frac{1}{2} (\mathcal{M}_{++,++} - \mathcal{M}_{+-,+-})  \,, \ \mathcal{M}_{TL}^{a} = \frac{1}{2} (\mathcal{M}_{++,00} - \mathcal{M}_{0+,-0}) \,.  
\label{eq:amp}
\end{gather}
Here, $q_i$ and $\lambda_i=0,\pm$ are respectively the momenta and helicities of the virtual photons.
The amplitudes are either even (first six) or odd (last two) with respect to the crossing-symmetric variable $\nu = q_1 \cdot q_2$. Using the optical theorem, one can relate the forward scattering amplitudes to two-photon fusion amplitudes $\mathcal{M}_{\lambda_1\lambda_2}$ associated with $\gamma^*(\lambda_1,q_1) \ \gamma^*(\lambda_2,q_2) \ \to X(p_X)$. More explicitly we~have
\begin{equation}
W_{\lambda_3 \lambda_4, \lambda_1 \lambda_2} = \mathrm{Im}\,\mathcal{M}_{\lambda_3 \lambda_4, \lambda_1 \lambda_2}  = \frac{1}{2} \int \mathrm{d} \Gamma_X (2\pi)^4 \delta(q_1+q_2-p_X)  \mathcal{M}_{\lambda_1 \lambda_2}(q_1,q_2,p_X) \, \mathcal{M}^{*}_{\lambda_3 \lambda_4}(q_1,q_2,p_X) \,.
\label{eq:opt}
\end{equation}
Using unitarity and analyticity, it is then possible to write dispersion relations at fixed values of the virtualities $Q_1^2$ and $Q_2^2$. Considering once-subtracted sum rules, they can be generically written as \cite{Pascalutsa:2012pr}
\begin{subequations}
\label{eq:sr}
\begin{align}
\mathcal{M}_{\rm even}(\nu)  &= \mathcal{M}_{\rm even}(0) + \frac{2 \nu^2}{\pi} \int_{\nu_0}^\infty \! d \nu^\prime \frac{1}{\nu^{\prime}(\nu^{\prime \, 2} - \nu^2-i \epsilon) } W_{\rm even}(\nu^\prime) \,, \\
\mathcal{M}_{\rm odd}(\nu)  &=  \nu \mathcal{M}^{\prime}_{\rm odd}(0) + \frac{2 \nu^3}{\pi} \int_{\nu_0}^\infty \! d \nu^\prime \frac{1}{ \nu^{\prime 2} ( \nu^{\prime \, 2} - \nu^2-i \epsilon) } W_{\rm odd}(\nu^\prime) \,,
\end{align}
\end{subequations}
where even/odd refers to the parity of the amplitude with respect to the variable $\nu$. The first step is to compute the eight forward amplitudes on the lattice. In a second step, we present a phenomenological model to describe the two-photon fusion processes and then evaluate the right-hand side of Eq.~(\ref{eq:sr}): the main ingredients are the single-meson TFFs. Finally, fitting the lattice data with our phenomenological model we obtain information about TFFs which, in turn, can be used as input parameter in the calculation of the HLbL contribution to the muon $(g-2)$.

\section{Lattice results}\label{sec-2}

\subsection{Lattice setup}

\begin{figure}[t]
\sidecaption
	\includegraphics*[width=0.40\linewidth]{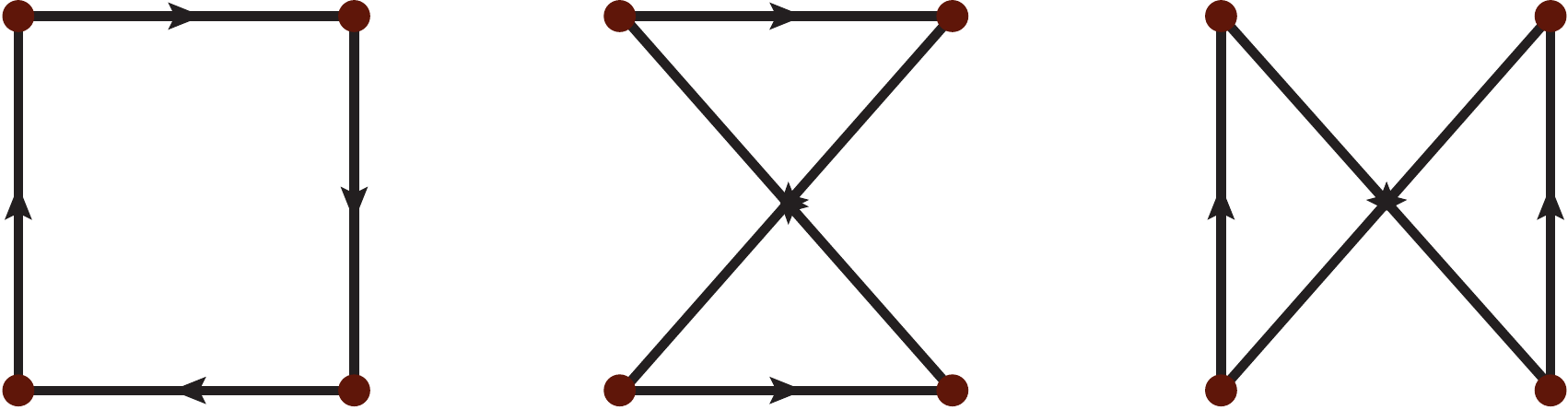}	
	\label{fig:4pt}
	\caption{Three of the six fully-connected quark contractions for four-point functions. The last three contractions are obtained after reversing the direction of the fermion flow}
\end{figure}

On the lattice, we compute the Euclidean four-point correlation function
\begin{equation*}
\Pi^E_{\mu\nu\rho\sigma}(Q_1,Q_2) = \sum_{X_1,X_2,X_4} 
\, \langle J^c_{\mu}(X_1)\, J^c_{\nu}(X_2)\, J^l_{\rho}(0)\, J^c_{\sigma}(X_4) \rangle_E \ e^{-i Q_1 (X_2-X_1)} \ e^{ i Q_2 X_4} \ + \ {\rm contact\ terms} \,,
\end{equation*}
where $J_{\mu}(x) = \frac{2}{3} \overline{u}(x)\gamma_{\mu}u(x) - \frac{1}{3} \overline{d}(x)\gamma_{\mu}d(x)$ is the electromagnetic current and the upper index $l,c$ refers to the local and point-split vector current respectively. The latter is exactly conserved on the lattice and does need renormalisation. For the local current, the renormalisation factor $Z_V^{{l}}$ has been computed non-perturbatively in Ref.~\cite{DellaMorte:2005xgj,Fritzsch:2012wq}. 
In this work, only the subset of fully connected diagrams, depicted in Fig.~(\ref{fig:4pt}), is considered.
We use five CLS (Coordinated Lattice Simulations) lattice ensembles with two degenerate dynamical quarks at two different lattice spacings  and pion masses down to $190~\MeV$. The ensembles, whose parameters are listed in Table~\ref{tabsim}, are generated using the plaquette gauge action for gluons~\cite{Wilson:1974sk} and the $\mathcal{O}(a)$-improved Wilson-Clover action for the fermions~\cite{Sheikholeslami:1985ij} with the non-perturbative parameter $c_{\rm SW}$~\cite{Jansen:1998mx}. For each ensemble, the four-point correlation function is computed at two values of $Q_1^2$ ($Q_1 = (0,0,0,1)$ and $Q_1 = (0,0,0,3)$ in units of $2\pi/T$ where $T$ is the time extent of the lattice) and for all values of $Q_2$ such that $Q^2_2 \lesssim 4~\GeV^2$.
The helicity amplitudes in Eq.~(\ref{eq:amp}) are finally obtained from the Euclidean four-point correlation function 
\begin{equation*}
\mathcal{M}(q^2_1,q^2_2,\nu) = e^4 \, T^E_{\mu\nu\mu'\nu'} \Pi^E_{\mu'\nu'\rho\sigma}(Q_1,Q_2) \,, \quad \mathcal{\overline{M}}(q^2_1,q^2_2,\nu) = \mathcal{M}(q^2_1,q^2_2,\nu) - \mathcal{M}(q^2_1,q^2_2,0) \,,
\end{equation*}
for some Euclidean tensor $T^E$. In particular, an explicit expression for $\mathcal{M}_{TT}$ can be found in~\cite{Green:2015mva}.

\begin{table}[b]
\small
\center
\caption{Parameters of the simulations: the bare coupling $\beta = 6/g_0^2$, the lattice size, the hopping parameter $\kappa$, the lattice spacing $a$ extracted from \cite{Fritzsch:2012wq}, the pion and rho masses and the number of gauge configurations.}
\begin{tabular}{lcl@{\hskip 02em}l@{\hskip 01em}c@{\hskip 01em}c@{\hskip 01em}c@{\hskip 01em}c@{\hskip 01em}c}
	\toprule
	CLS	&	$\quad\beta\quad$	&	$L^3\times T$ 		&	$\kappa$		&	$a~[\fm]$	&	$m_{\pi}~[\MeV]$	&	$m_{\rho}~[\MeV]$	& $m_{\pi}L$	 &	$\#$~confs\\
	\midrule
	E5	&		$5.3$		&	$32^3\times64$	& 	$0.13625$	& 	$0.0652(6)$  	& 	$437(4)$ &	971	&	 4.7 	&	500\\  
	F6	&		 	 		& 	$48^3\times96$	&	$0.13635$	& 				& 	$314(3)$ &	886	&	 5.0	&	125\\      
	F7	&		 	 		& 	$48^3\times96$	&	$0.13638$	& 				& 	$270(3)$ &	841	&	 4.3 	&	150\\       
	G8	&		 	 		& 	$64^3\times128$	&	$0.136417$	& 				& 	$194(2)$ &	781	&	 4.1	&	124\\         
	\midrule
	N6	&		 $5.5$ 		& 	$48^3\times96$	&	$0.13667$	& 	0.0483(4)		& 	$342(3)$ &	917	&	 4.0	&	86\\        
	\bottomrule
 \end{tabular} 
\label{tabsim}
\end{table}

\subsection{Connected contribution to the light-by-light forward scattering amplitudes}

\begin{figure}[t]
	
	\center
	\includegraphics*[width=0.49\linewidth]{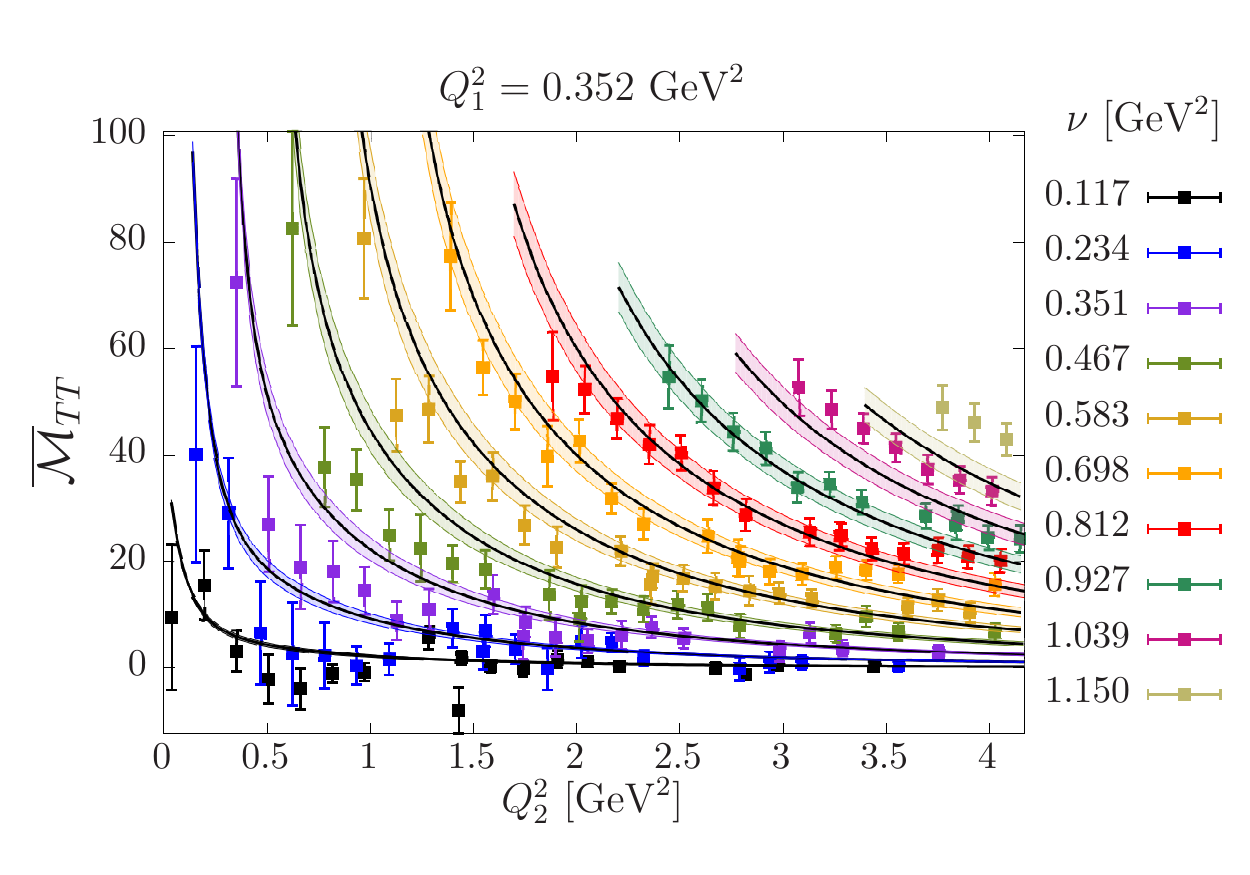}
	\includegraphics*[width=0.49\linewidth]{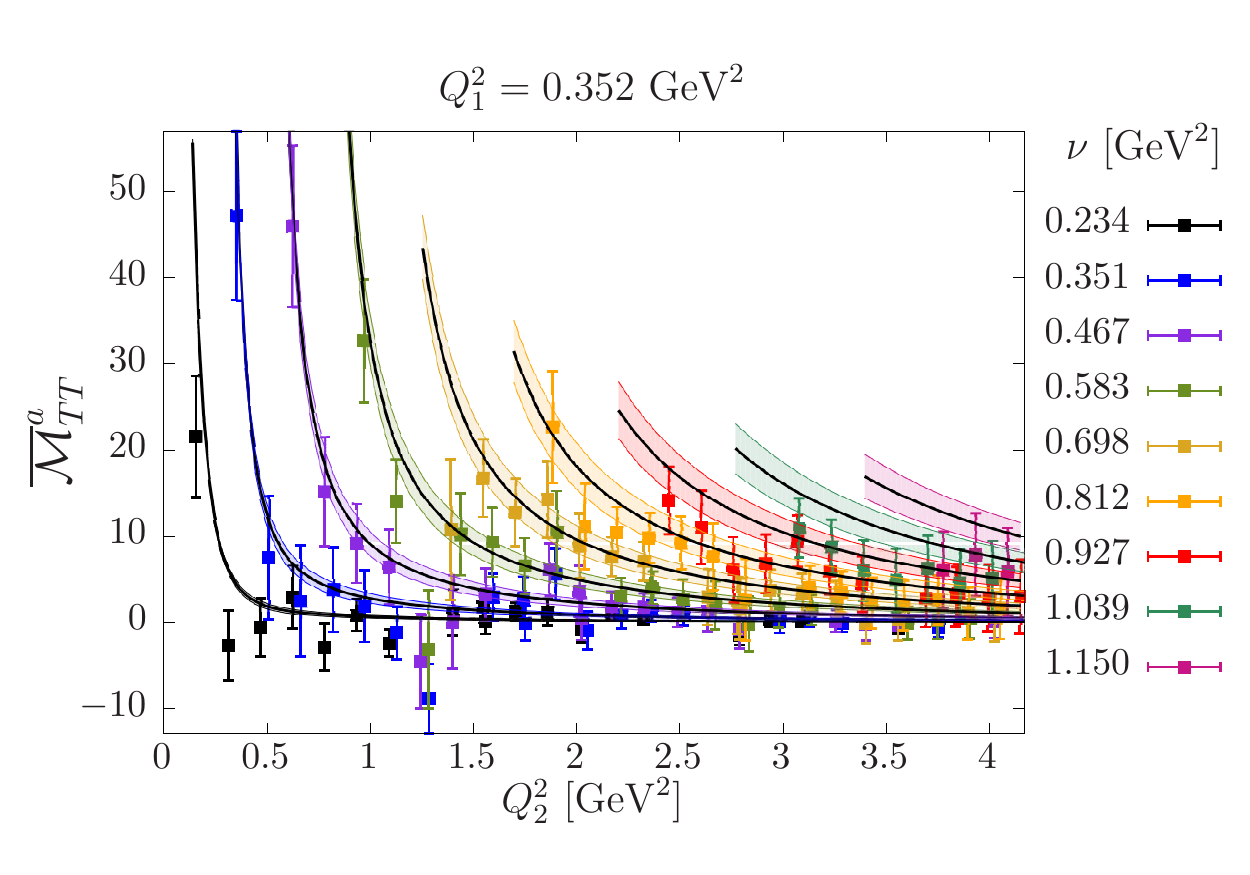} \\ \vspace{-0.5cm}
	\includegraphics*[width=0.49\linewidth]{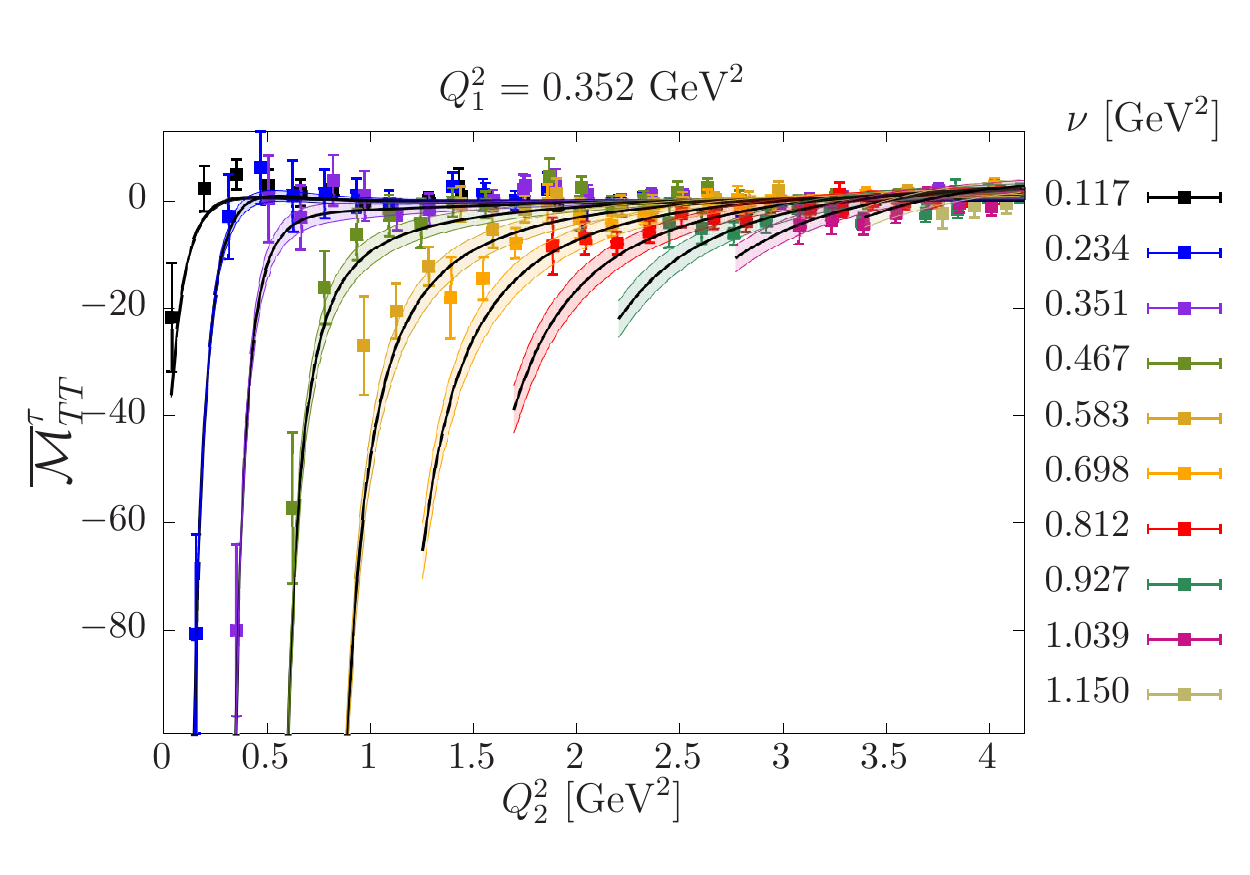}
	\includegraphics*[width=0.49\linewidth]{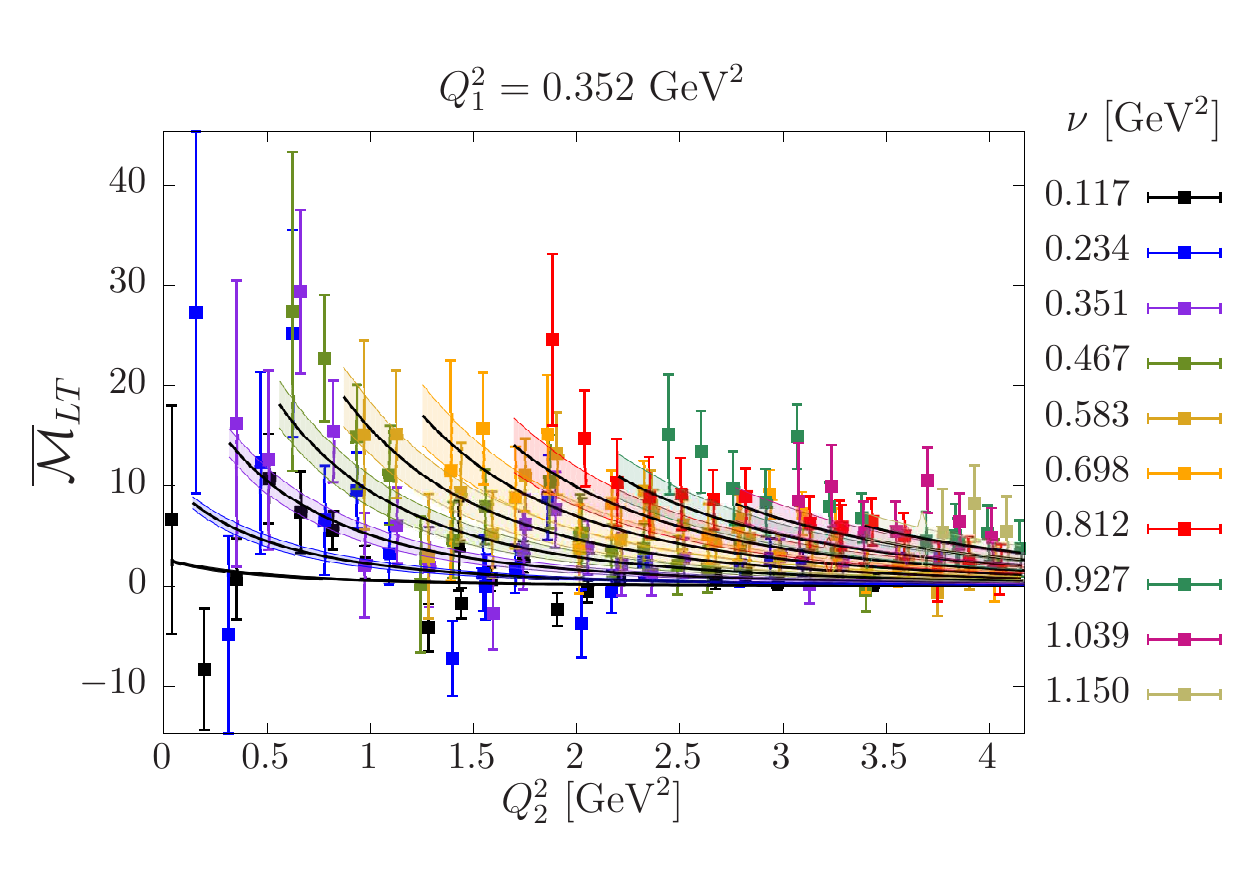} \vspace{-0.5cm}

	\caption{Amplitudes $\MsTT$, $\MsTTa$, $\MsTTt$ and $\MsLT$ for the ensemble F7 with $Q_1^2=0.352~\GeV^2$. The curves with error-bands represent the result of the fit discussed in Sec.~\ref{sec:fits}.  \label{fig:amp}}
\end{figure}

The result for a subset of the eight amplitudes for the ensemble F7 is depicted in Fig.~\ref{fig:amp} for a fixed virtuality $Q_1^2 = 0.352~\GeV^2$. The amplitude $\MsTT$ is positive as it corresponds to a physical cross section whereas $\MsTTa$, $\MsTTt$ and $\MsLT$ are related to interference terms in Eq.~(\ref{eq:opt}) and are not sign definite. 

\section{Description of lattice data using phenomenology}\label{sec:model}

\subsection{Parametrization of the hadronic two-photon fusion cross sections}

In this section, we describe the main assumptions used to describe our data using phenomenology.
Concerning single-meson production, only C-parity-even states are involved. We therefore include pseudoscalar, scalar, axial and tensor mesons and consider only the lightest state in each channel. Since we are working with two degenerate dynamical quarks, isoscalar $\eta$-type mesons are not included. Furthermore, based on flavor symmetry and large-N counting (N is the number of colors)~\cite{Bijnens:2016hgx}, one can show that with two dynamical quarks, the non-singlet meson poles contribute with a factor $34/9$ to the fully-connected diagrams and the singlet mesons do not contribute whereas in the 2+2 disconnected diagrams, in addition to the singlet mesons, the non-singlet mesons contribute with a factor $-25/9$ thus largely compensating for the previous factor 34/9. Therefore, since we neglect disconnected contributions, we do not include isoscalar and isovector mesons separately but only isovector mesons with an overall factor $34/9$, and the list of particles is given in Table~\ref{tab:particles}.
Finally, the Born approximation to $\gamma^*\gamma^* \to \pi^+\pi^{-}$ cross section is included using scalar QED as in Ref.~\cite{Pascalutsa:2012pr} dressed with a monopole vector form factor. Since this contribution turns out to be small, the monopole mass will not be fitted but set to the lattice rho mass. 
\begin{table}[t] 
\small
\center
\begin{center} 
\caption{Mass and two-photon width of the particles included in this work as quoted by the PDG~\cite{Olive:2016xmw}. For the axial meson, we use the isoscalar two-photon width divided by 25/9 based on isospin and large-$N$ arguments.} \label{tab:particles}
\begin{tabular}{l@{\hskip 0.2in}lcc@{\hskip 0.2in}lcc@{\hskip 0.2in}lcc}
\toprule  
&	$\pi$ ($0^{-+}$)		&	$a_0(980)$ ($0^{++}$)	&	$a_1(1260)$ ($1^{++}$)	&	$a_2(1320)$ ($2^{++}$)	\\
\midrule
$m$~[MeV]	&	134.98	&	$980(20)$		&	$1230(40)$	&	$1318.3(0.6)$ \\
\midrule 
$\GammaGG$~[keV] 	&	$0.0078(5)$ 	&	$0.30(10)$ 	&	$\emph{1.26}$  &	$1.00(6)$	  \\
\bottomrule
\end{tabular} 
\end{center}
\end{table} 
Our simulations are performed away from the physical quark masses; therefore, the pion and rho meson masses are set to their lattice values determined from the exponential decay of the pseudoscalar and vector two-point correlation functions. For other resonances, we simply assume a constant shift in the spectrum, such that $m_X = m_X^{\phys} + (m_{\rho}^{\lat} - m_{\rho}^{\exp})$.

The pion TFF has been measured experimentally by several collaborations in the single-virtual case for virtualities in the range $Q^2 \in [0.6-40]~\GeV^2$~\cite{Behrend:1990sr,Gronberg:1997fj,Aubert:2009mc,Uehara:2012ag}. The experimental data are well described by a monopole TFF but the latter is ruled out by lattice data obtained with both single and double virtual photons~\cite{Gerardin:2016cqj}. We therefore use the lattice results obtained on the same set of ensembles.

The scalar meson can be produced by two transverse~(T) or two longitudinal photons~(L) and the two-photon fusion process is therefore parametrized by two TFFs $F^T_{{\cal S} \gamma^\ast \gamma^\ast}$ and $F^L_{{\cal S} \gamma^\ast \gamma^\ast}$. The transverse TFF $F^T_{{\cal S} \gamma^\ast \gamma^\ast}$ for the isoscalar meson $f_0(980)$ has been measured experimentally in the region $Q^2~<~30~\GeV^2$ by the Belle Collaboration~\cite{Masuda:2015yoh} and the results are compatible with a monopole form factor with $M_S = 800(50)~\MeV$. We therefore assume 
\begin{equation*}
\frac{  F^T_{{\cal S} \gamma^\ast \gamma^\ast}(Q_1^2, Q_2^2) }{ F^T_{{\cal S} \gamma^\ast \gamma^\ast}(0, 0)  } = \frac{1}{( 1 + Q_1^2/M_S^2)(1 + Q_2^2/M_S^2)} \,,
\end{equation*}
where $M_S$ will be considered as a free fit parameter. The normalisation of the TFF is given by the two-photon decay width $\Gamma_{\gamma\gamma}$~\cite{Olive:2016xmw} and we assume that $F^L_{{\cal S} \gamma^\ast \gamma^\ast}(Q_1^2, Q_2^2)$ and $ F^T_{{\cal S} \gamma^\ast \gamma^\ast}(Q_1^2, Q_2^2)$ are equal.

For axial mesons, the two-photon fusion cross section is parametrized by two form factors $F^{(0)}_{{\cal A} \gamma^\ast \gamma^\ast}$ and $F^{(1)}_{{\cal A} \gamma^\ast \gamma^\ast}$ where $\Lambda=0,1$ corresponds to the two helicity states of the axial meson. Inspired by quark models, we use the same parametrization as in Ref.~\cite{Pascalutsa:2012pr}
\begin{align*}
F^{(0)}_{{\cal A} \gamma^\ast \gamma^\ast}(Q_1^2, Q_2^2) &= m_A^2 A(Q_1^2, Q_2^2) \,, \\
F^{(1)}_{{\cal A} \gamma^\ast \gamma^\ast}(Q_1^2, Q_2^2) &= - \frac{\nu}{X} \left( \nu + Q_2^2\right) \, m_A^2 A(Q_1^2, Q_2^2) \,, \\
F^{(1)}_{{\cal A} \gamma^\ast \gamma^\ast}(Q_2^2, Q_1^2) &= - \frac{\nu}{X} \left(\nu + Q_1^2\right) \, m_A^2 A(Q_1^2, Q_2^2) \,,
\end{align*}
in which $2\nu = m_A^2 + Q_1^2 + Q_2^2$ with $m_A$ the meson mass and $A(Q_1^2, 0) / A(0, 0)   = 1/ (1 + Q_1^2/M_A^2)^2$ and assuming factorisation such that $A(Q_1^2, Q_2^2) = A(Q_2^2, Q_1^2)$.
The L3 Collaboration has measured the TFF for the isoscalar meson $f_1(1285)$ in the single-virtual case in the region $Q^2 < 5~\GeV^2$~\cite{Achard:2001uu,Achard:2007hm}, and their result reads $M_A = 1040(78)~\MeV$. In the following, $M_A$ will be considered as a free fit parameter and the normalisation of the TFF is given from the effective two-photon width as defined in~Ref.~\cite{Pascalutsa:2012pr}.

Finally, tensor meson amplitudes are described by four form factors $F^{(\Lambda)}_{{\cal T} \gamma^\ast \gamma^\ast}(Q_1^2, Q_2^2)$ with $\Lambda = (0,T),\ (0,L),\ 1,\ 2$.
The single-virtual TFFs with helicities $\Lambda=(0,T),1,2$ have also been measured experimentally in the region $Q^2 < 30~\GeV^2$ by the Belle Collaboration~\cite{Masuda:2015yoh}, and data are compatible with a dipole form factor~\cite{Danilkin:2016hnh}
\begin{equation*}
\frac{  F^{(\Lambda)}_{{\cal T} \gamma^\ast \gamma^\ast}(Q_1^2, Q_2^2) }{ F^{(\Lambda)}_{{\cal T} \gamma^\ast \gamma^\ast}(0, 0)  }  =   \frac{1}{( 1 + Q_1^2/M_{T,(\Lambda)}^2)^2(1 + Q_2^2/M_{T,(\Lambda)}^2)^2} \,.
\end{equation*}
The four dipole masses $M_{T,(\Lambda)}$ will be considered as free fit parameters. Again, the normalisation of the TFFs are obtained from the experimentally measured two-photon width~\cite{Olive:2016xmw} assuming that the ratio of helicity two to helicity zero mesons is $r = 91.3~\%$ \cite{Dai:2014lza}. For helicity $(0,L)$, where no experimental data exist, we used the value extracted from Ref.~\cite{Danilkin:2016hnh} using dispersive sum rules.

\subsection{Fits of the eight light-by-light forward scattering amplitudes and chiral extrapolations}\label{sec:fits}

\begin{figure}[h!]

	\center
	\includegraphics*[width=0.44\linewidth]{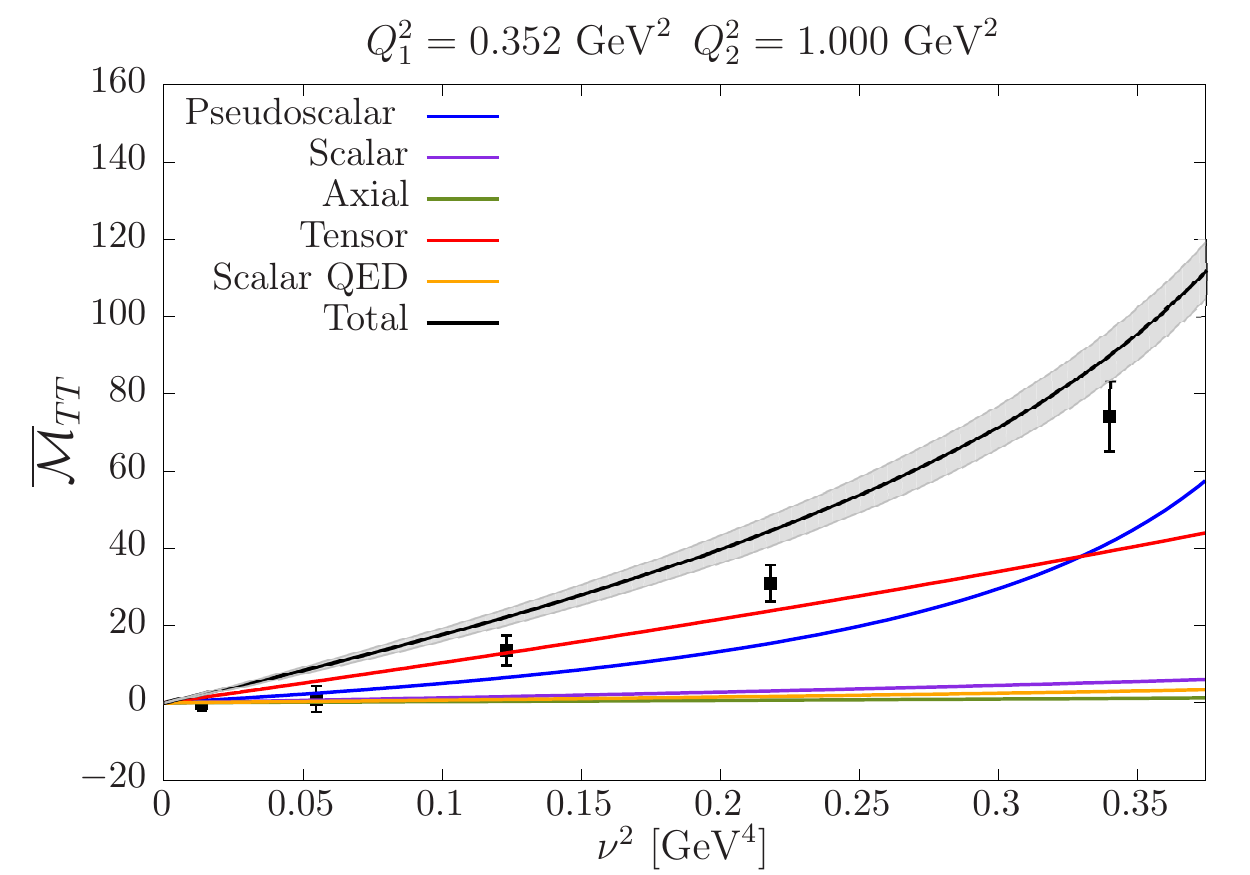}
	\includegraphics*[width=0.44\linewidth]{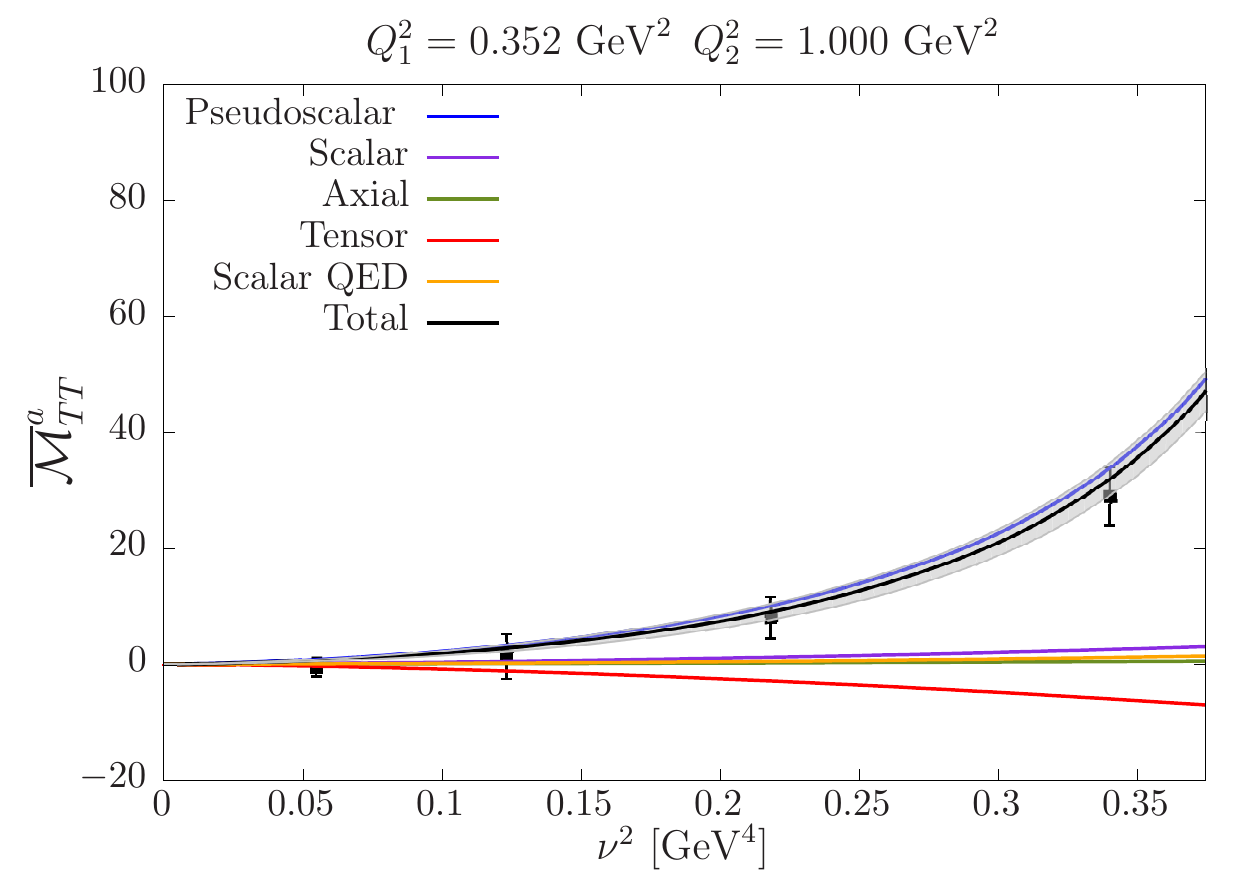} \\
	\includegraphics*[width=0.44\linewidth]{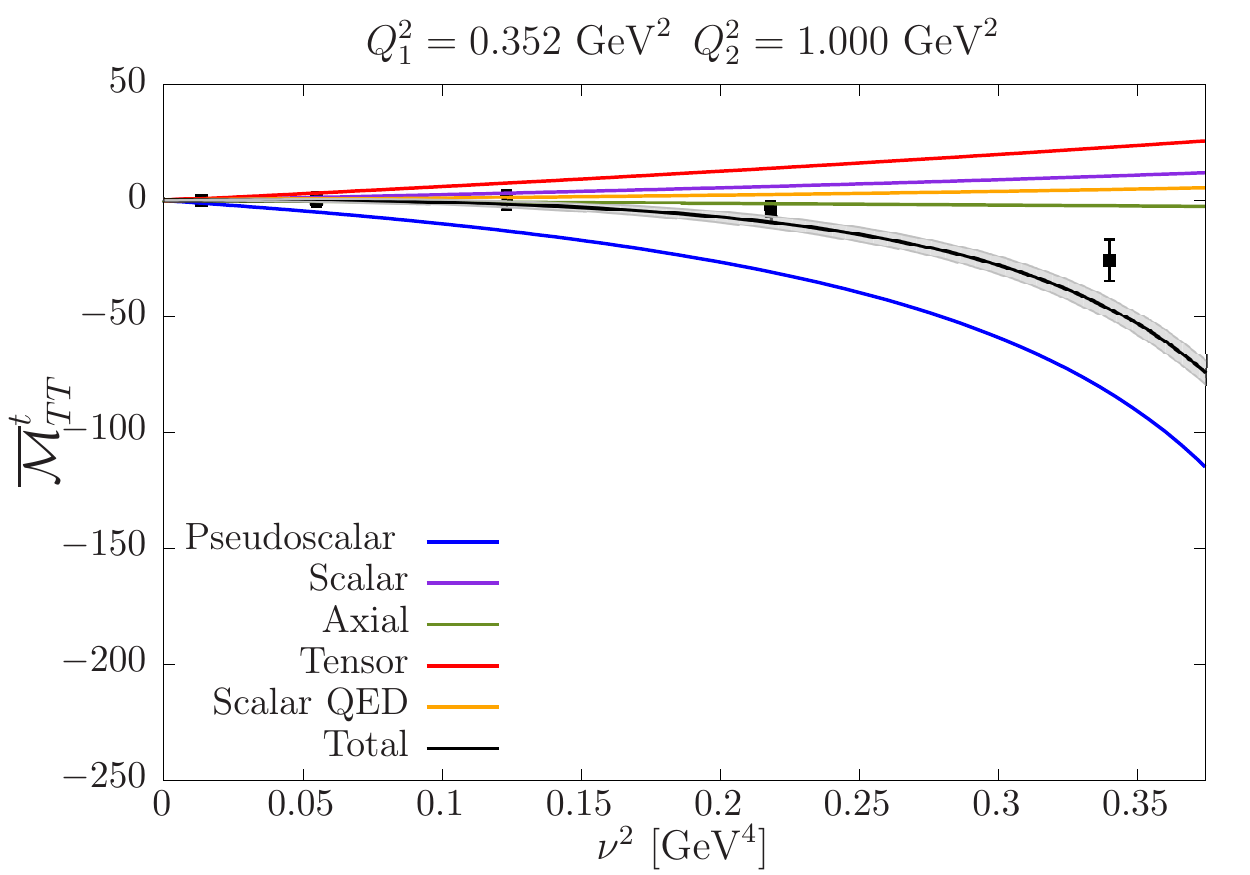}
	\includegraphics*[width=0.44\linewidth]{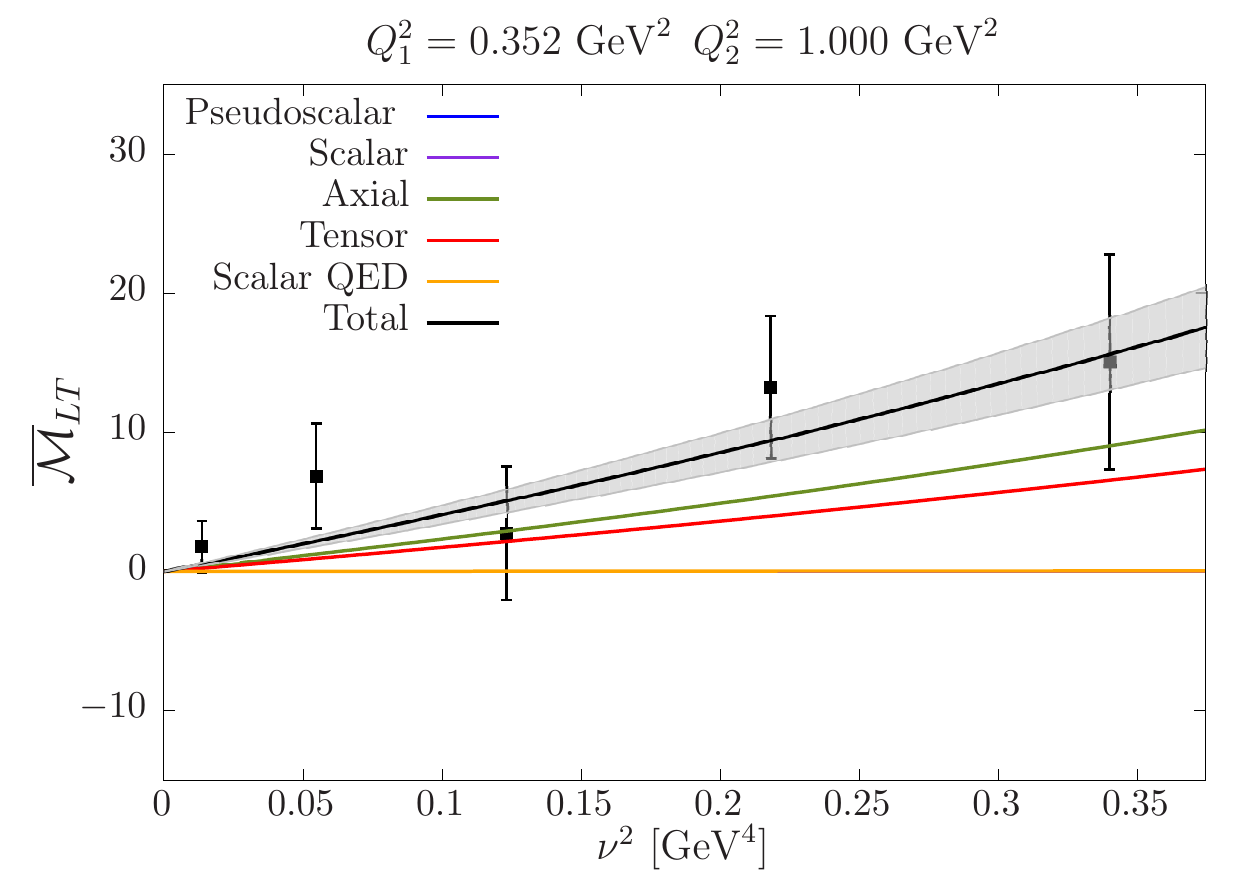} \vspace{-0.3cm}

	\caption{The dependance of each amplitudes on $\nu$ for two different values of $Q_2^2$ and with $Q_1^2=0.377 \GeV^2$. Results correspond to the ensemble F7. The black line corresponds to the total contribution and coloured lines include only single-meson contribution. \label{fig:nu_dep} }
	\vspace{-0.6cm}
\end{figure}

We perform a global fit of the eight amplitudes using the phenomenological model described in the previous section. There are six fit parameters corresponding to monopole and dipole masses of the scalar, axial and tensor mesons. The results for four amplitudes corresponding to the ensemble F7 are shown in Fig.~\ref{fig:amp} where the $\chi^2/\dof$ is 1.2, and the results for all five ensembles are summarized in Table~\ref{tab:fit}.
In Fig.~\ref{fig:nu_dep}, we show the relative contribution of each channel to the different amplitudes at fixed virtualities $Q_1^2 = 0.352~\GeV^2$ and $Q_2^2 = 1.000~\GeV^2$. In Table~\ref{tab:contrib} we also give the relative contribution of each channel to the eight amplitudes.
\begin{table}[thb]
\small
\centering
\caption{Results of the fit of the eight subtracted amplitudes $\MsTT$, $\MsTTt$, $\MsTTa$,  $\MsTL$, $\MsLT$, $\MsTLa$, $\MsTLt$ and $\MsLL$ for the five lattice ensembles.}
\begin{tabular}{l@{\hskip 01em}c@{\hskip 01em}c@{\hskip 01em}c@{\hskip 01em}c@{\hskip 01em}c@{\hskip 01em}c@{\hskip 01em}c}
	\toprule
		&	$M_{S}~[\GeV]$	&	$M_{A}~[\GeV]$	&	$M^{(2)}_{T}~[\GeV]$	&	$M^{(0,T)}_{T}~[\GeV]$ &	$M^{(1)}_{T}~[\GeV]$	& $M^{(0,L)}_{T}~[\GeV]$ & $\chi^2/\dof$ \\
	\midrule
	E5  	&	1.38(11)	&	1.26(10)	&	1.93(3)	&	2.24(5)	 &	2.36(4)	&	0.60(10)	& 4.2	\\
	F6  	&	1.12(14)	&	1.44(5)	&	1.66(9)	&	2.17(5)	 &	1.85(14)	&	0.89(28)	& 1.2	\\
	F7  	&	1.04(18)	&	1.29(8)	&	1.61(12)	&	2.08(7)	 &	2.03(7)	&	0.57(16)	& 1.2	\\
	G8  	&	1.07(10)	&	1.36(5)	&	1.37(24)	&	2.03(6)	 &	1.63(13)	&	0.73(14)	& 1.1	\\
	N6  	&	0.86(37)	&	1.59(3)	&	1.72(17)	&	2.19(4)	 &	1.72(18)	&	0.51(8)	& 1.4	\\
	\bottomrule
 \end{tabular} 
\label{tab:fit}
\end{table}
\begin{table}[t!]
\small
\center
\caption{Relative contributions in \% of each particle to the different amplitudes for the ensemble F7 using $Q_1^2 = 0.352~\GeV^2$, $Q_2^2 = 1~\GeV^2$, $\nu = 0.467~\GeV^2$. }
\begin{tabular}{l@{\hskip 02em}c@{\hskip 02em}c@{\hskip 02em}c@{\hskip 02em}c@{\hskip 02em}c@{\hskip 02em}c@{\hskip 02em}c@{\hskip 02em}c}
	\toprule
			&	$\MsTT$	&	$\MsTTa$	&	$\MsTTt$	&	$\MsTL$	&	$\MsLT$	&	$\MsTLa$	&	$\MsTLt$	&	$\MsLL$	 	 \\
	\midrule
	$0^{-+}$  		&	35	&	68	&	$-56$	&	$\times$	&	$\times$	&	$\times$	&	$\times$	&	$\times$ \\
	$0^{++}$		&	7	&	8	&	11	&	$\times$	&	$\times$	&	23		&	14	&	42	\\
	$1^{++}$		&	2	&	1	&	$-2$	&	43		&	57		&	$-43$		&	32	&	$\times$\\
	$2^{++}$		&	53	&	$-20$	&	25	&	56		&	42		&	19		&     $-47$	&	25\\
	Scalar QED	&	4	&	3	&	5	&	1		&	$<1$		&	$-15$		& 	$-7$	&	33	\\
	\bottomrule
 \end{tabular} 
\label{tab:contrib}
\vspace{-0.2cm}
\end{table}

The pseudoscalar and tensor mesons give the dominant contribution to the amplitudes $\MsTT$, $\MsTTt$ and $\MsTTa$ involving two transverse photons. Here, the axial meson gives a small contribution, especially at low virtualities as expected from the Landau-Yang theorem.
The pseudoscalar does not contribute to $\MsTL$, $\MsLT$ where the main contribution comes from axial and tensor mesons. In the last three amplitudes $\MsTLa$, $\MsTLt$, $\MsLL$ all scalar, axial and tensor contribute significantly, sometimes with opposite signs. 
Finally, scalar QED contribution is always small compared to other channels.

\begin{figure}[thb]

\centering
	\includegraphics*[width=0.44\linewidth]{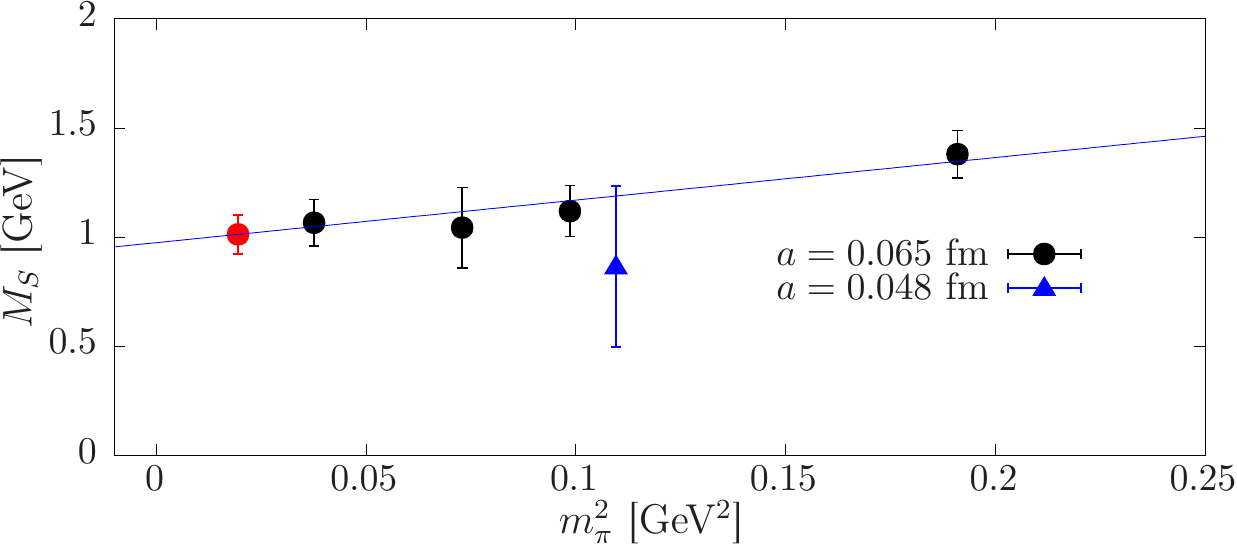}
	\includegraphics*[width=0.44\linewidth]{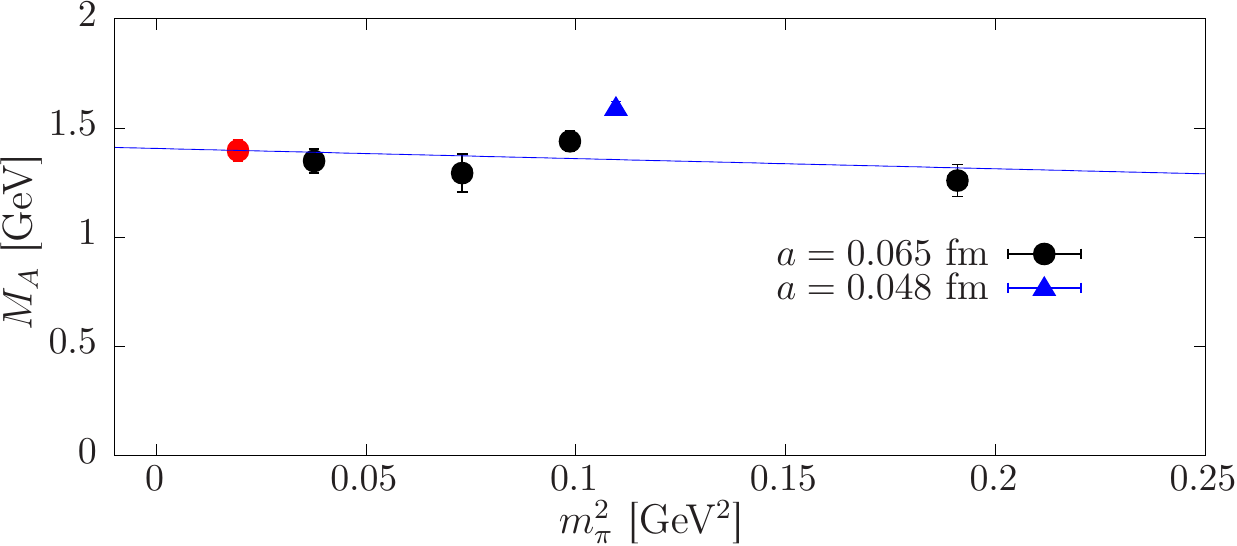}\\
	\includegraphics*[width=0.44\linewidth]{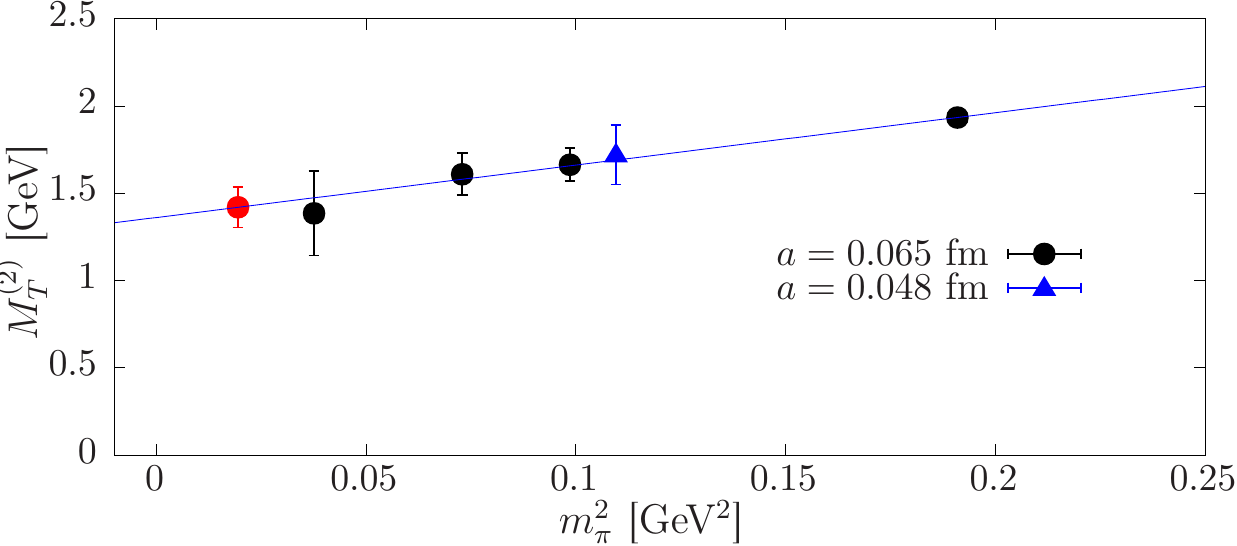}
	\includegraphics*[width=0.44\linewidth]{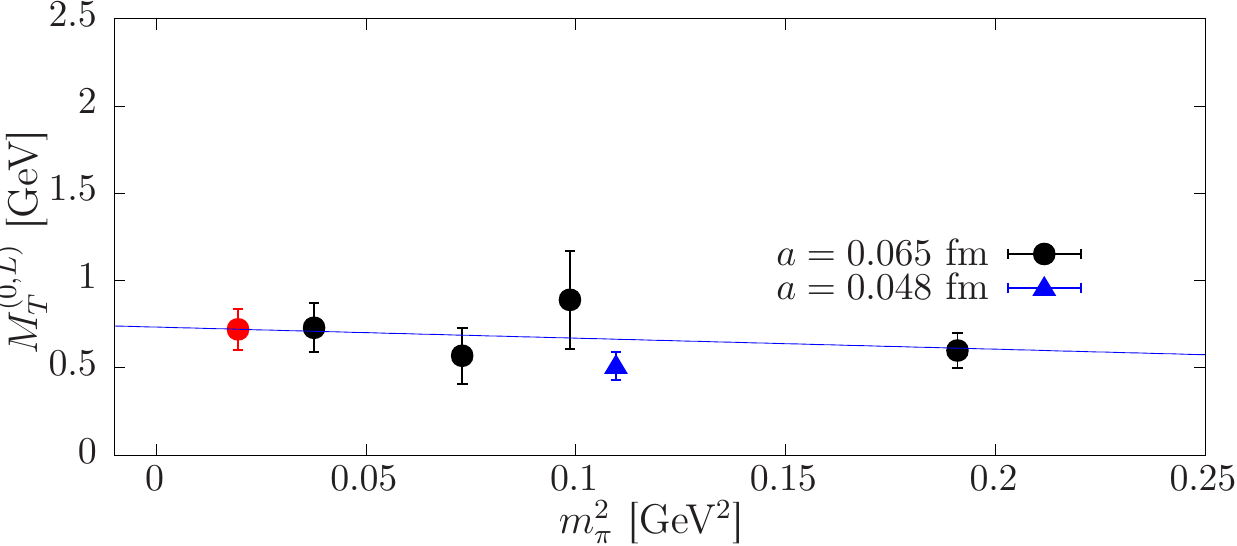}\\	
	\includegraphics*[width=0.44\linewidth]{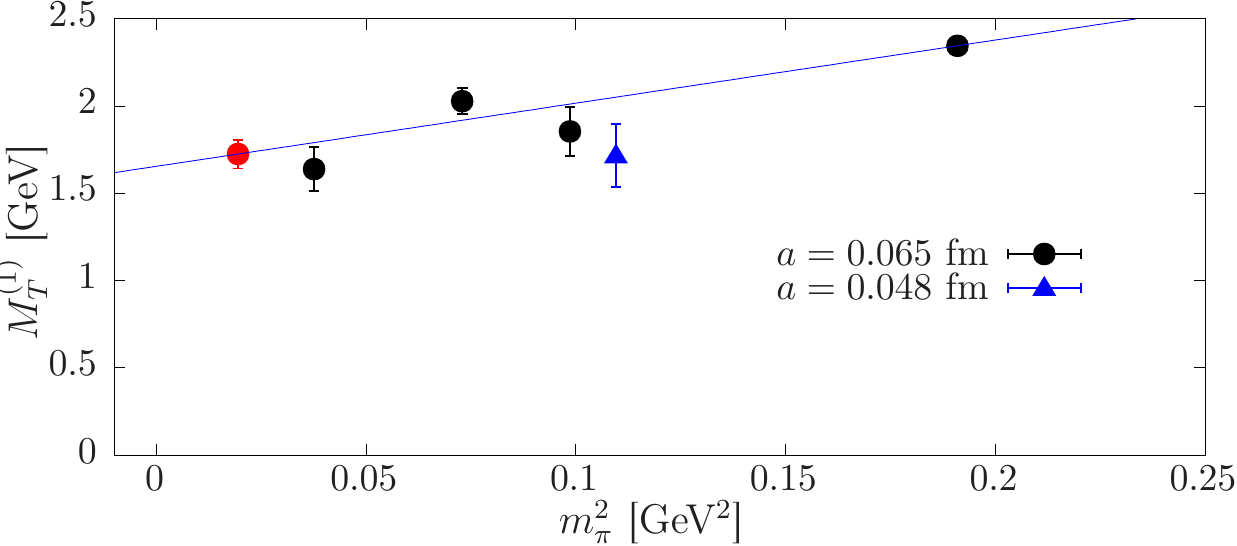}
	\includegraphics*[width=0.44\linewidth]{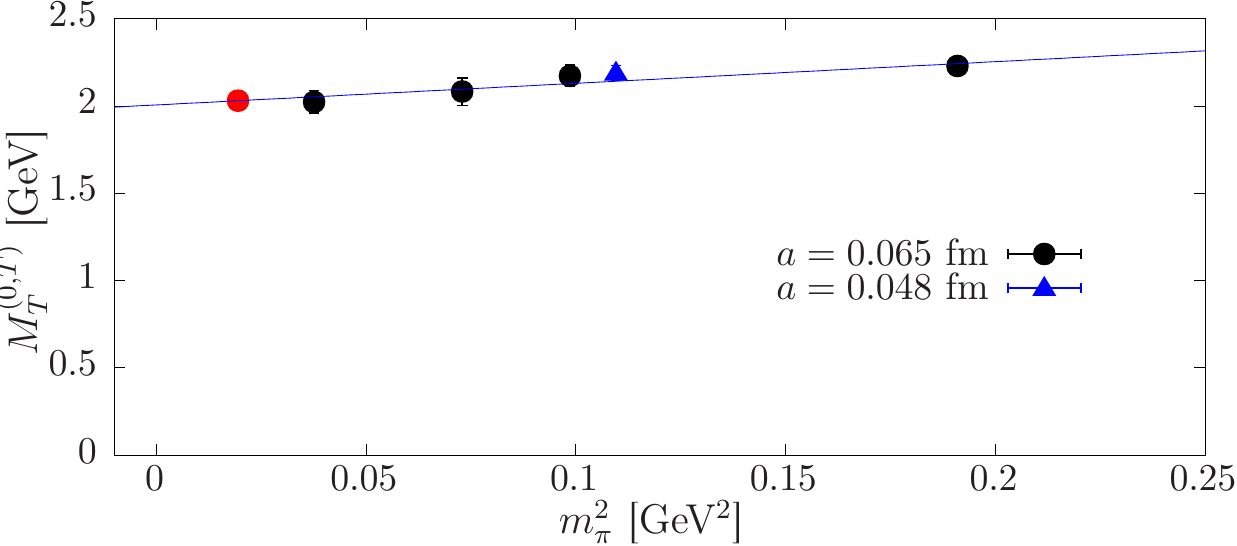}
	
	\vspace{-0.3cm}
	\caption{Chiral extrapolations of the scalar, axial and tensor monopole/dipole masses. The black points correspond to lattice data, the red point to the result extrapolated at the physical pion mass and the blue point corresponds to a smaller lattice spacing and gives an indication of the size of discretization effects. \label{fig:extrap} }
\vspace{-0.9cm}
\end{figure}

The chiral extrapolation is depicted in Fig.~\ref{fig:extrap}. The scalar monopole mass is $M_S =~1.01(9)~\GeV$, slightly above the experimental result from the Belle Collaboration, which obtains $M_S = 796(54)~\MeV$ for the isoscalar scalar meson~\cite{Masuda:2015yoh}. The axial dipole mass $M_A = 1.40(5)~\GeV$ should be compared with the experimental value by the L3 Collaboration $M_A = 1040(80)~\MeV$ for the isoscalar $f_1(1285)$~\cite{Achard:2001uu,Achard:2007hm}. 
Finally, the tensor dipole mass $M^{(2)}_{T} = 1.42(12)~\GeV$ is close to the value obtained in \cite{Danilkin:2016hnh} from a fit to the Belle data~\cite{Masuda:2015yoh}. However, the dipole masses $M^{(1)}_{T} = 1.73(10)~\GeV$ and $M^{(0,T)}_{T} = 2.03(6)~\GeV$ are above the values quoted in~\cite{Danilkin:2016hnh}  ($M^{(1)}_{T} = 0.916(20)~\GeV$, $M^{(0,T)}_{T} = 1.051(36)~\GeV$).

\section{Conclusion}\label{sec:conclusion}

We have presented a first lattice calculation of the eight light-by-light forward scattering amplitudes. Only the fully-connected subset of diagrams is included so far and we plan to include the 2+2 disconnected diagrams which have been computed for the ensembles E5 and F6. The amplitudes satisfy dispersion relations which involve the two-photon fusion processes. Assuming that only a few states are needed to saturate the sum rules, we have proposed a phenomenological model to describe our data. The main inputs are the associated single meson TFFs. For the pion, we use our recent lattice calculation and for scalar, axial and tensor mesons, we assume monopole and dipole TFFs, where monopole and dipole masses are considered as free fit parameters whereas the normalisation is taken from experiment. The model is in good agreement with lattice data and using different pion masses we are able to extrapolate our results to the chiral limit and compare with experiment. \\

\vspace{-0.2cm}
\begin{acknowledgement}
We acknowledge the use of computing time on the JUGENE and JUQUEEN
machines located at FZ J\"ulich, Germany. The four-point functions were
computed on the cluster ``Clover'' of the Helmholtz-Institute Mainz and the ``Mogon'' cluster at the University of Mainz. The programs were written using QDP++~\cite{Edwards:2004sx} with the deflated SAP+GCR solver from openQCD~\cite{OpenQCD}.
This work was supported by the DFG through SFB1044.
\end{acknowledgement}

\vspace{-0.2cm}
\bibliography{latbib}

\end{document}